\documentclass[11pt]{article}
\usepackage{moriond2000,epsfig}

\def\Journal#1#2#3#4{{#1} {\bf #2}, #3 (#4)}

\def\be{\begin{equation}}
\def\ee{\end{equation}}
\def\bea{\begin{eqnarray}}
\def\eea{\end{eqnarray}}

\begin{document}
\vspace*{4cm}
\title{THE LENS-REDSHIFT TEST REVISITED}

\author{ P. HELBIG }

\address{Rijksuniversiteit Groningen, Kapteyn Instituut, Postbus 800,\\
NL-9700 AV Groningen, The Netherlands}

\maketitle\abstracts{
Kochanek\cite{K92} suggested that the redshifts of gravitational lens
galaxies rule out a large cosmological constant.  This result was
questioned by Helbig \& Kayser\cite{HK}, who pointed out that selection
effects related to the brightness of the lens can bias the results of
this test against a high $\lambda_{0}$ value; however, we did not claim
that the observations \emph{favoured} a high $\lambda_{0}$ value, merely
that current observational data were not sufficient to say either way,
using the test as proposed by Kochanek\cite{K92} but corrected for
selection effects.  Kochanek\cite{K96a} pointed out that an additional
observable, namely, the fraction of measured lens redshifts, provides
additional information which restores the sensitivity of the test to the
cosmological model, at least somewhat.  Here, I consider three aspects. 
First, I discuss the appropriateness of the correction to the test
proposed by Kochanek (1996a). Second, I briefly mention the slightly
different statistical methods which have been used in connection with
this test.  Third, I discuss what results can be obtained today now that
more and better-defined observations are available. 
}

\section{Introduction}

The optical depth for gravitational lensing depends on the cosmological
model, the Faber-Jackson and Tully-Fisher relations, lens-galaxy type
(or the morphological mix), the luminosity function of lens galaxies and
the $S$-$z$ relation of the source population (e.g.\
Kochanek\cite{K92}, Helbig \& Kayser\cite{HK}).  There is an
obvious problem with simply measuring the integrated optical depth,
i.e.\ the number of lens systems (according to some useful definition): 
There is a degeneracy between various parameters such that quite
different combinations can result in the same number of lenses.  While
it is possible to break this degeneracy somewhat, this requires a
careful survey and cannot be done with a sample of lenses `from the
literature'.  Kochanek\cite{K92} pointed out that one could use the 
\emph{shape} of the optical-depth function $\mathrm{d}\tau/\mathrm{d}z$ 
as a probe of the cosmological model.  The advantage of this approach is 
that it does not depend on the overall normalisation, as counting the 
number of lenses obviously does.  Also, it is quite sensitive to the 
cosmological model, with the dependence on the cosmological model of a) 
the combination of angular size distances and b) the volume element, 
both of which appear in $\mathrm{d}\tau/\mathrm{d}z$, reinforcing one 
another.  In other words, the redshifts of lens galaxies can be used as 
a probe of the cosmological model which is relatively little affected by 
our ignorance of other factors which determine the total optical depth.

\section{History}

Kochanek\cite{K92} used a sample of 4 gravitational lens systems from
the literature (estimating the lens redshift from absorption lines if
unknown) and found that the Einstein-de~Sitter model was 5--10 times
more likely than a flat model dominated by a cosmological constant.
Helbig \& Kayser\cite{HK} pointed out that this is potentially
subject to a strong bias:  It could be that most known lens redshifts are
low not because we live in a universe in which this is more probable,
but since we could not have measured them if they were higher.  To
correct for this effect, we suggested comparing the shape of
$\mathrm{d}\tau/\mathrm{d}z$ not over the whole range
[0,$z_{\mathrm{s}}$] (in practice, the value of this function is
negligible before $z_{\mathrm{s}}$ is reached), but rather only out to
that redshift where a lens redshift could have been measured, assuming
some realistic limiting magnitude (at this redshift,
$\mathrm{d}\tau/\mathrm{d}z$ usually still has a non-negligible value) and
found that no interesting constraints could be obtained from
then-current data (using 6 systems, all with measured, not estimated,
lens redshifts), even if many more such systems were found, and that 
this conclusion did not depend on the precise value assumed for the 
limiting magnitude.

Kochanek\cite{K96a} then pointed out that one can use an additional
observable to restore cosmological sensitivity to the lens-redshift
test: the fraction of lens systems with measured redshifts.  If a strong 
bias were present such that only low lens redshifts could be measured, 
then there should be many lens systems with unmeasured redshifts.  While 
true, this misses the point of Helbig \& Kayser\cite{HK}:  Our claim 
was not that the observations supported a large value of the 
cosmological constant (nor the opposite), but rather that the conclusion 
of Kochanek\cite{K92} did not follow from the sample used (or our 
sample) since the lens-brightness bias had not been taken into account.
Also, the correction proposed in Kochanek\cite{K92} assumes that 
unknown lens redshifts are unknown only because they are faint; in
practice, there can be many other reasons why some lens redshifts have
not yet been measured (e.g. the maximum declination accessible from 
UKIRT). 

Various different statistical measures have been used to compare the
observed and predicted lens-redshift distributions.  Here, I only
consider the maximum-likelihood method (e.g.\ Kochanek\cite{K96a}),
which I consider to be most appropriate.  However, results from using
the binning method of Helbig \& Kayser\cite{HK} or a Kolmogorov-Smirnov
test (Helbig, unpublished) give qualitatively similar results. 

\section{Using CLASS}

The whole issue of unknown lens redshifts and their possible causes can 
be avoided if one has a sample which is complete with respect to lens 
redshifts.  CLASS (e.g. Helbig\cite{CLASS}) is close to this goal, and 
the JVAS subset of CLASS (more exactly, the JVAS lens systems in CLASS 
which are also part of the statistically complete lens-survey sample; 
see Helbig\cite{CLASS} for more details) is actually complete.  While 
only consisting of four systems, this is the same number used in 
Kochanek\cite{K92}, so the time is ripe to revisit this topic.  (The 
last JVAS lens redshift was obtained by Kochanek \& Tonry\cite{TK}.)

\begin{figure}
\psfig{figure=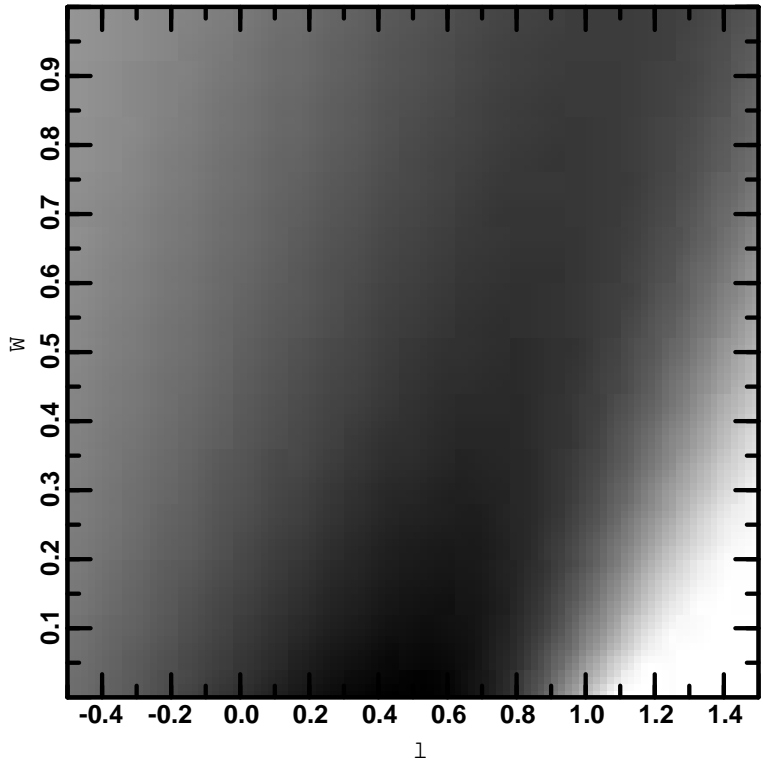,height=4in} 
\caption{Likelihood as a function of $\lambda_{0}$ and $\Omega_{0}$ 
using the Kochanek sample; darker means higher likelihood.
\label{fig:kochanek}}
\end{figure}

\begin{figure}
\psfig{figure=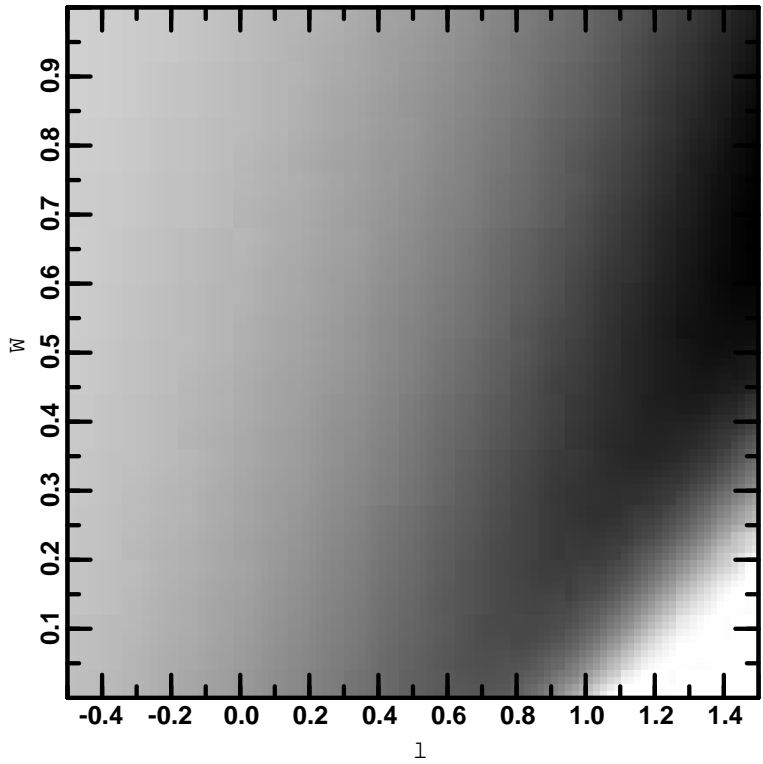,height=4in} 
\caption{Likelihood as a function of $\lambda_{0}$ and $\Omega_{0}$ 
using the JVAS sample; darker means higher likelihood.
\label{fig:JVAS}}
\end{figure}

Figure~\ref{fig:kochanek} shows the likelihood as a function of 
$\lambda_{0}$ and
$\Omega_{0}$ for the sample from Kochanek\cite{K92} 
while Fig.~\ref{fig:JVAS} shows
the same for the JVAS lens systems B0218+357, MG0414+054, B1030+074 and
B1422+231.  
It is obvious that the Kochanek\cite{K92} sample indicates
that the Einstein-de~Sitter model is more likely than a
flat model dominated by a cosmological constant.  The JVAS sample tells 
a different story.  Probably, part of the difference, in particular, the 
low probability of models near the white area to the lower right
(which corresponds to no-big-bang models and is excluded \textit{a 
priori}) can be explained by the bias noted in 
Helbig \& Kayser\cite{HK}, while part can be explained by small-number 
statistics.  This will be explored in more detail in Helbig \& 
Rusin\cite{HR}.  (It should be noted that the results for the 
Kochanek\cite{K92} sample presented here do not correspond exactly to 
those in Kochanek\cite{K92} since there (as in Helbig \&
Kayser\cite{HK}), the now-known-to-be-erroneous $(3/2)^{\frac{1}{2}}$ 
factor for elliptical galaxies was used.  Including this factor increases the 
relative likelihood of the Einstein-de~Sitter model for the Kochanek\cite{K92} 
sample while its effect on the JVAS sample is less pronounced.)

\section{Conclusions and Future Prospects} 

It is obvious that the conclusion of Kochanek\cite{K92} was premature:
using a better-defined and in particular bias-free (since complete)
sample, the lens-redshift test does not disfavour cosmological-constant
dominated models, although the significance of this is not yet clear.
Since the publication of Kochanek\cite{K92}, of course, the
cosmological constant has become popular again and, although more
detailed lens-statistics analyses are not incompatible with this (e.g.\
Helbig\cite{Helbig}), it is not yet clear whether systematic effects,
such as our lack of sufficient information about the $S$-$z$ plane of
the source population (e.g. Kochanek\cite{K96b}), make current estimates
of $\lambda_{0}$ from the analysis of lens surveys unreliable.  It is at
least interesting that the lens-redshift test does not seem to favour an
Einstein-de~Sitter universe over a model (flat or not) dominated by a
cosmological constant.  When the much larger CLASS sample is complete
with respect to lens redshifts, the time will be ripe to revisit this
topic once again. 

\section*{Acknowledgments}

It is a pleasure to thank David Rusin for useful discussions and the 
CLASS team for providing the numbers to work with.


\section*{References}
\begin{thebibliography}{99}

\bibitem{K92}C.S.~Kochanek, \Journal{ApJ}{384}{1}{1992}

\bibitem{HK}P.~Helbig \& R.~Kayser, \Journal{A\&A}{308}{359}{1996}

\bibitem{K96a}C.S.~Kochanek, \Journal{ApJ}{466}{638}{1996}

\bibitem{CLASS}P.~Helbig, these proceedings

\bibitem{TK}J.L.~Tonry \& C.S.~Kochanek, \Journal{AJ}{117}{2034}{1999}

\bibitem{HR}P.~Helbig \& D.~Rusin, in preparation

\bibitem{Helbig}P.~Helbig \Journal{A\&A}{350}{1}{1999}

\bibitem{K96b}C.S.~Kochanek, \Journal{ApJ}{473}{595}{1996}

\end{thebibliography}
\end{document}